\documentclass[sort&compress
    ,final            
,numberedheadings 
  ]
  {aipproc}

\layoutstyle{6x9}

\def\apgt{\ {\raise-.5ex\hbox{$\buildrel>\over\sim$}}\ }
\def\aplt{\ {\raise-.5ex\hbox{$\buildrel<\over\sim$}}\ }

\newcommand{\rs}{\mbox {$R_{\odot}$}}
\newcommand{\ms}{\mbox {$M_{\odot}$}}
\newcommand{\myr}{\mbox {~${\rm M_{\odot}~yr^{-1}}$}}
\def\path#1{(path~{\bf #1})}
\newcommand{\mc}{\mbox {$M_{\rm Ch}$}}
\newcommand{\pyr}{\mbox {{\rm yr$^{-1}$}}}
\newcommand{\nusn}{\mbox {$\nu_{\mathrm{Ia}}$}}

\begin{document}

\title{Population synthesis for low and intermediate mass binaries}

\author{L. R. Yungelson}{
  address={Institute of Astronomy, 48 Pyatnitskaya Str., 119017, Moscow, Russia}
}

\begin{abstract}
A review of the basic principles of population synthesis for binary stars is presented. We discuss the break-up of low and intermediate mass close binaries over different evolutionary scenarios and, as an example, briefly consider results of the population synthesis for  SN\,Ia. 
\end{abstract}

\maketitle

\section{INTRODUCTION}
Population synthesis for binary stars is a
convolution of the
statistical data on initial parameters and birthrates of  binaries
with the  evolutionary scenarios for them.
The main goals of population synthesis are:

$\bullet$
to understand the descent of binaries of different types from unevolved main-sequence  binaries which
differ only in the masses of components $M_1, M_2,$ and their separations $a$ and to understand the links between the objects of different types;

$\bullet$
to estimate the incidence of binaries of different
types and occurrence rates of certain {\sl events} in them (for instance,
explosions of Novae or Supernovae);

$\bullet$
to  find the distributions of binaries of different
types over parameters like masses of components, orbital
periods, luminosity of components, spatial velocities;

$\bullet$
to model `observed'' samples of stars applying selection effects.

Below, we review the basic principles of population synthesis, discuss the break-up of low and intermediate mass close binaries over different evolutionary scenarios and, as an example, consider results of population synthesis studies for Supernovae Ia (SN\,Ia). We define  \emph{low and intermediate mass binaries as objects in which components end their evolution as white dwarfs}.   

\section{EVOLUTIONARY SCENARIOS FOR BINARY STARS}   

The main concept of the population synthesis is that of an ``evolutionary scenario'', 
 the sequence of transformations of a
binary system with given initial $(M_{10}, M_{20}, a_0$) that it can experience in  Hubble time (the term is coined by van den Heuvel and Heise \cite{heuvelheise72}). 

\begin{figure}
 \label{fig:seprad}
  \includegraphics[scale=0.4,angle=-90]{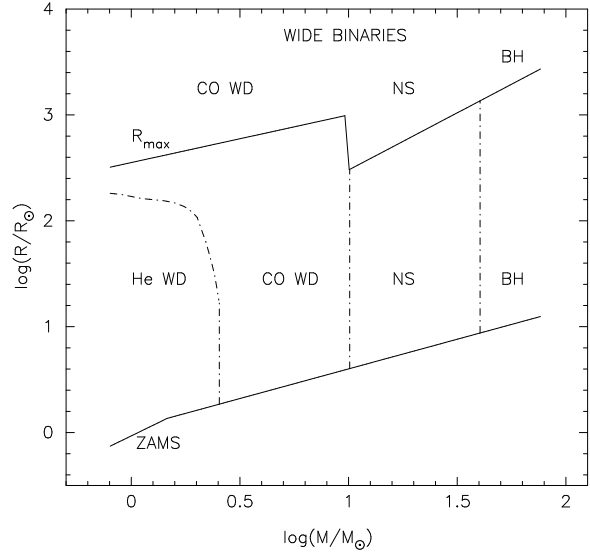}
   \caption{Descendants of components of close binaries depending on the
radius of star at the instant of RLOF.
Close in evolutionary sense stars fill Roche lobes before expanding to
$R_{\rm max}$. For wide binaries the products of evolution are the same as for
single stars.
   }
\end{figure}

For construction of  evolutionary scenarios all binaries are separated into ``close'' and ``wide'' ones. In close binaries components interact between themselves, the principal form of interaction is mass exchange through Roche lobe overflow. The possibility for interaction of components in a given binary is defined by  critical radius $R_{cr}$, a boundary within which the atmosphere of the evolving star can expand while star remains a distinctly separate object. If $R_\star > R_{cr}$, stellar matter starts to flow out.
Figure~\ref{fig:seprad} sketches ZAMS- and maximum radii of stars and shows descendants of  components of binaries depending on their mass and radius at the instant of RLOF (for solar chemical composition stars). The upper mass limit for
progenitors of He white dwarfs (WD) is certain to several tenth of \ms. The
boundary  between progenitors of WD and neutron stars (NS) is known to 
about 1\ms\ (it may be different for single stars and components of close 
binaries). The least certain is the lower cut-off of the masses
of progenitors of black holes (BH). It is expected that between 20 and 50\,\ms, components of close binaries may produce NS or BH, depending on such
parameters
as stellar winds, rotation, and magnetic fields \cite{eh98a,wl99,fry99}; the limit of 40\,\ms\ is sketched in Fig.~1
provisionally.  

The data necessary for construction of evolutionary scenarios is provided by stellar evolution computations. The main data are: the rates of mass loss through RLOF or stellar wind (SW)  for stars in different evolutionary stages and the nature of the products of evolution; time scales of evolutionary stages; initial--final mass relations; 
response of 
stars to accretion;
transformations of separations due to mass exchange/loss and angular momentum loss, especially in the common envelope stages. 

Evolutionary scenarios for  binaries with certain initial parameters may be convolved with
star formation rate, 
binarity rate, distributions of binaries over $M_{10}$, $a_0$, $q_0=M_{20}/M_{10}$  to get birthrates of stars of different types. Convolved with lifetimes this gives
number distribution of stars of different kinds at any epoch of the Galactic history, their distributions over basic parameters, and occurrence rates of different events, e. g., SN. Applying  selection effects one gets  ``observed samples'' of different binaries.
The first numerical population synthesis studies were carried out independently by Kornilov and Lipunov \cite{kornilovlipunov83} for NS,
Dewey and Cordes \cite{dc87} for radiopulsars,
Politano \cite{pol88} for cataclysmic variables,
Lipunov and Postnov \cite{lippost88} for compact magnetized stars.

Binarity rate for stars is not known for sure, 
it was studied for separate groups of stars only. 
Studies of selection effects for spectroscopic and visual binaries \cite{pty82,vtykp88} suggest that binarity rate may be close to 100\%.

The studies of spectroscopic and visual binaries \cite{pty82,vtykp88} 
suggest that (i) IMF for primary components of binaries may be well approximated by power law $dn/dM \propto M^{-2.5}$;
(ii) distribution of binaries is flat in $\log a$. This allows to estimate that among binaries $\sim 40\%$ are close and
$\sim 96\%$ are low and intermediate mass ones.

\subsection{Common envelopes}

The crucial problem that plagues all population synthesis studies is uncertainty about mass and angular momentum loss (AML) from the system and especially about \emph{common envelopes} (CE). It is assumed that 
CE arise when accretor in a close binary is unable to accrete all matter supplied by the donor \cite{pac76}.
The usual assumption is that the  envelope of the donor engulfs
both components; the latter spiral in, losing energy. This results in reduction  of the separation of components and may
end either in
the merger of components or in the dissipation of CE.
This is understood qualitatively, but  neither analytical theory nor convincing enough
computations of the processes inside common envelopes exist. Therefore, usual approach  is based on comparison of the energy released in  rapprochement of components and binding energy of CE
 by applying (most commonly) an equation \cite{web84,dek90}
\begin{equation}
 \label{eq:ce}
  \frac {M_{\rm i}(M_{\rm i}-M_{\rm f})}{ \lambda R} =
  \eta_{\rm ce}  \left( \frac {M_{\rm f}m}{2a_{\rm f}} - 
  \frac{M_{\rm i}m}{2a_{\rm i}}\right), 
\end{equation}
where indexes $i$ and $f$ refer to initial and final masses and separations, $m$ is the mass of accretor,
$\eta_{\rm ce}$ is the efficiency of deposition of energy into common envelope, 
$\lambda$ is a function of stellar radius that, in essence, describes how well donor envelope is bound to the core (it varies $\sim3 - 4$ times along RG and  AGB and depends on the definition of ``core'' boundary and possible release of thermal energy \cite{hpe94,dt00,td01}. If whole ensemble of close binaries is considered and Eq.~(\ref{eq:ce}) is applied to any system, irrespective to the evolutionary state of components, satisfactory agreement with observations is usually achieved if
$\eta_{\rm ce}\lambda \simeq  1 - 2$.

However,  Nelemans and co-authors \cite{nvy+00}, for instance,
have shown that it is impossible to reconstruct evolution of observed close binary helium WD
back to main-sequence binaries if Eq.~(\ref{eq:ce}) 
is applied for the stage of unstable mass transfer between components of comparable mass.
Instead,  it was suggested in \cite{nvy+00} to use for this stage an equation based on the angular momentum ($J$) balance: 
\begin{equation}
 \label{eq:dj}
  \frac {\Delta J}{J} = \gamma  \left( \frac {\Delta M} {M_d + M_a} \right),
\end{equation}
where $\Delta M$ is the amount of matter lost by the system. 
\citet{nelemans04} have shown that applying   Eq.(~\ref{eq:dj}) with $\gamma\simeq1.5$ it is possible to explain the origin of all known binary WD and pre-CV.
Equations (\ref{eq:ce}) and (\ref{eq:dj}) give similar results if masses of components strongly differ \cite{nelemans04}.

In the case of super-Eddington accretion in the systems with NS or BH accretors formation of common envelopes, perhaps, may be avoided.
The 
energy of the accreted matter
released close to the compact object (at $r_{\rm c}$) may be  sufficient to expel the ``excessive'' matter
from the Roche-lobe surface around the compact object (at $r_{\rm L,c}$) if accretion rate 
$\dot M \leq \dot M_{\rm Edd}\,r_{\rm L,c}/r_{\rm c}$ (see, e. g., \cite{kb99}). It is also suggested that  common envelopes
 may be
avoided
by  driving excess of matter by optically-thick winds from the surface of hydrogen-burning accreting white dwarfs \cite{kathac94}. Note,  that the model of radiatively driven winds still lacks  elaborated theoretical justification, but it is supported by the presence of outflows in many systems with compact components.

If CE formation is avoided 
as described above, it is usually \emph{assumed} that
the matter leaves the system with
specific angular momentum of the accreting component. This \emph{assumption} proved to be useful, for instance,  for the models of formation of millisecond radiopulsars \cite{ts99,ths00} or ultracompact low-mass X-ray binaries with WD donors  \cite{ynh02}.   

\subsection{Evolutionary paths of low and intermediate mass close binaries}  

\begin{figure}
 \label{fig:close_ce}
  \caption{Flowchart of the evolution for close binaries that experience unstable 
first RLOF. The numbers indicate ``path'' discussed in the text. For some important channels the fraction of all close binaries passing through them is indicated. 
  }
   \includegraphics[scale=0.5]{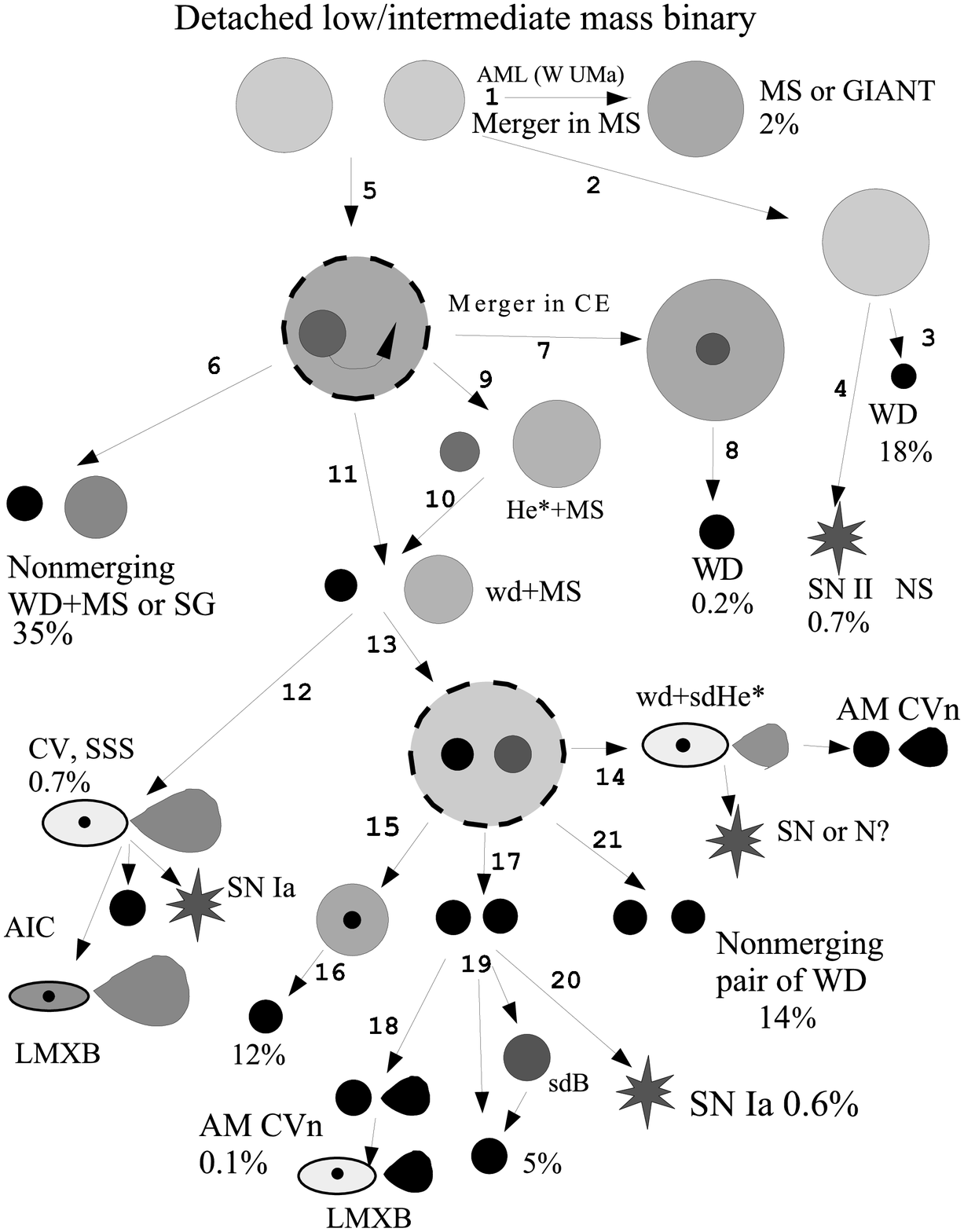}
\end{figure}

\begin{figure}[t!]
 \label{fig:rlof}
  \includegraphics[scale=0.5]{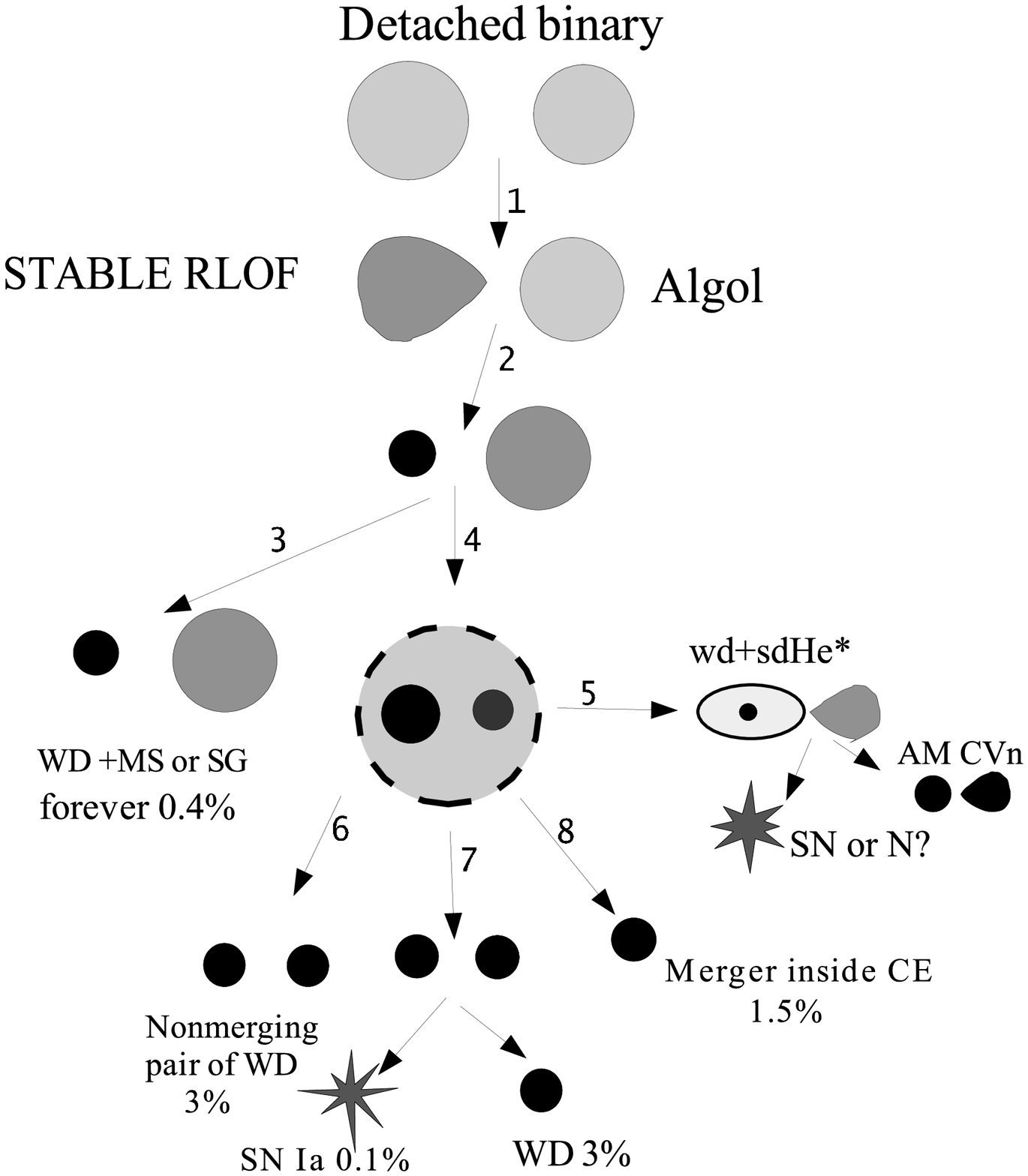}
   \caption{Flowchart of the evolution for close binaries that experience stable 
first RLOF.}

\end{figure}

The character of mass-exchange in a binary depends on the mass ratio of components and the structure of the envelope of Roche-lobe filling star \cite{tfy82,hw87}. For stars with radiative envelopes, to the first approximation,  mass exchange is stable if $q \aplt 1.2$; for $1.2 \aplt q \aplt 2$ it proceeds in the thermal time scale of the donor; for $q \apgt 2$ it proceeds in the dynamical time scale. Mass loss occurs in dynamical time scale if the donor has deep convective envelope. In practice this means that overwhelming majority ($\sim 90\,\%$) of close binaries pass through one to four stages of common envelope.   

A run of population synthesis code (e. g., \cite{ty02}) for $\sim 1.5\times10^6$  close binaries with $0.8 \leq M_{10}/\ms \leq 11.4$  produces $\sim 600$ different scenarios of evolution.   
In Fig.~\ref{fig:close_ce} we present 
a kind of summary of scenarios for systems that have first unstable RLOF.
For description, we denote
evolutionary routes in Fig.~\ref{fig:close_ce} as ``path {\bf N}''.

In the tightest systems primaries overflow Roche lobe 
when 
they are 
still in MS. The least massive of these systems  ($M_{10} \aplt 1.5\,\ms$) evolve under the influence of AML via magnetically coupled stellar wind (MSW) and gravitational waves radiation (GWR). An outcome of evolution may be formation of a W\,UMa system that merges into rapidly rotating single star 
\path{1}. About 2\% of these systems do not finish their evolution in Hubble time
and remain MS-stars or giants. Majority of merger products end their evolution as WD \path{2--3}. Peculiarly, some merger products accumulate mass $\apgt 10\,\ms$ and explode as SN\,II \path{4}.

In wider systems donors overflow Roche lobes when they have He- or CO-cores. About one third of binaries avoid merger in CE and form pairs consisting of a He- or CO/ONe-WD and a low-mass companion that does not evolve past MS or subgiant stage. The systems remain wide enough to avoid merger due to AML via GWR and/or MSW \path{6}. Systems that merge in CE typically finish their evolution as WD \path{7--8}.

If a binary with a He-core donor does not merge in CE, it first forms a He-star+MS-star pair \path{9}, that may be observed as an sdB star with MS companion.    When He-star finishes its evolution, a pair harboring a CO white dwarf and a MS-star appears \path{10}. Similar systems are produced by binaries where RLOF occurs at AGB \path{11}. 

The tightest of the latter systems ($a \simeq$ several \rs) that have low-mass ($\aplt 1.5\,\ms$) MS-companions to WD evolve under AML and may form cataclysmic variables \path{12}. If accreted hydrogen
burns at the surface of WD stably, the system may appear as supersoft X-ray source (SSS in the plot). The outcome of the evolution of CV is not completely clear.
The donor may be disrupted when its mass decreases below
several hundredth of \ms\ \cite{rs83}; accretor may accumulate enough mass to explode as SN\,Ia (see below); WD may experience an accretion induced collapse into a NS and a low-mass X-ray binary may be formed. Actually, for a considerable fraction of CV evolution is, most probably, frozen
when $\dot{M}$ becomes low.

The second common envelope may form when MS-companion to WD overfills its Roche-lobe \path{13}. If the donor has a He-core and system does not merge, an sdB+WD system may emerge \path{14}. If separation of components is 
sufficiently small, AML via GWR may bring He-star into contact while He is still burning in its core. If $M_{\rm He}/M_{\rm wd} \aplt 1.2$ stable mass exchange is possible with typical $\dot M \sim 10^{-8}$\,\myr \cite{skh86}.
 It was suggested that accretion of He onto CO WD at such $\dot M$ leads to SN,
  when accreted He-layer detonates and initiates detonation of C (edge-lit detonation \cite{lg91}). However, it seems more likely that the lifting effect of rotation reduces the power of He-ignition and a He-nova is produced instead \cite{yoon_langer04}. If the system survives SN or nova explosion, an AM CVn-type system may be formed.

Some 12\% of close binaries merge in the second CE
and form single 
WD \path{15--16}. This may create certain excess of massive WD. If the system avoids merger, a pair of WD is formed \path{17}. The closest of them may be brought into contact by AML via GWR. If in a CO+He white dwarf pair conditions for stable mass exchange are fulfilled, an AM CVn system forms \path{18}, see for details \cite{ty96,npv+01,marsh_nel_st04}. 
About 5\% of close binary WD merge \path{19}; it is not yet clear how the merger proceeds and it is possible that for He+He or CO+He pairs a helium star is an intermediate stage
\citep{ guerrero04}.

A fraction of 
 close binaries produces merging CO WD pairs (``double-degenerates, DD'') with total mass greater than Chandrasekhar mass \mc \path{20} that may explode as SN\,Ia \cite{ty81,web84,it84a}. They 
are probably the most 
promising
candidates for SN\,Ia (see below).

Figure \ref{fig:rlof} shows typical evolutionary scenarios for the systems that stably exchange mass in the first RLOF. It applies to $\sim 10\%$ of all close binaries. In the first RLOF \path{1} binaries may be identified with Algol-type systems. Algol phase ends by formation of WD+MS-star system directly \path{3} or with an interfering phase of He-star+MS-star (similarly to \path{9-10} in Fig.~\ref{fig:close_ce}). In rare cases WD+MS remain ``frozen'' in this state \path{3}. RLOF by the former secondary usually results in CE, since $q \gg 1$
\path{4}. If the system did not merge in CE, the donor had a nondegenerate He-core and system is close enough, a semidetached WD+He-star system may arise due to the AML via GWR and in such a system a SN 
or He-nova may explode \path{5}. In this case formation of an AM CVn star also is possible, like in \path{14} in Fig.~\ref{fig:close_ce}. 
For the systems that avoid merger in CE
the following outcomes of evolution become possible: formation of a non-merging in a Hubble time pair of WD \path{6}; formation of a DD that merges due to AML via GWR and possibly produces SN\,Ia or a massive WD \path{7}. Merger in CE also produces a massive WD \path{8}.

A special case of evolution, that involves binaries with both low/intermediate and high mass components, leads to formation of recycled pulsars. First, in the system with initial $q_0\sim10$ the primary becomes a NS. If the system is not disrupted by SN explosion, low-mass secondary after a stage of  stable RLOF (when the system is a low-mass X-ray source) becomes a He WD. In another scenario $q_0$ is larger and, before forming a WD companion to NS, the system passes through CE.
In both cases NS are recycled by accretion.

Another peculiar example of evolution involves binaries with $q_0 \sim 1$ and mass of components  $\apgt (6 - 7)\,\ms$. If the first mass exchange is stable, initially  less massive component accumulates enough mass to explode as SN II. In this case massive white dwarf (former primary) gets a young neutron star companion (see \cite{py99,ts00} for details).    

\section{ CANDIDATE~ SUPERNOVAE Ia}

\begin{figure}[t!]
 \label{fig:snflow}
  \includegraphics[scale=0.5]{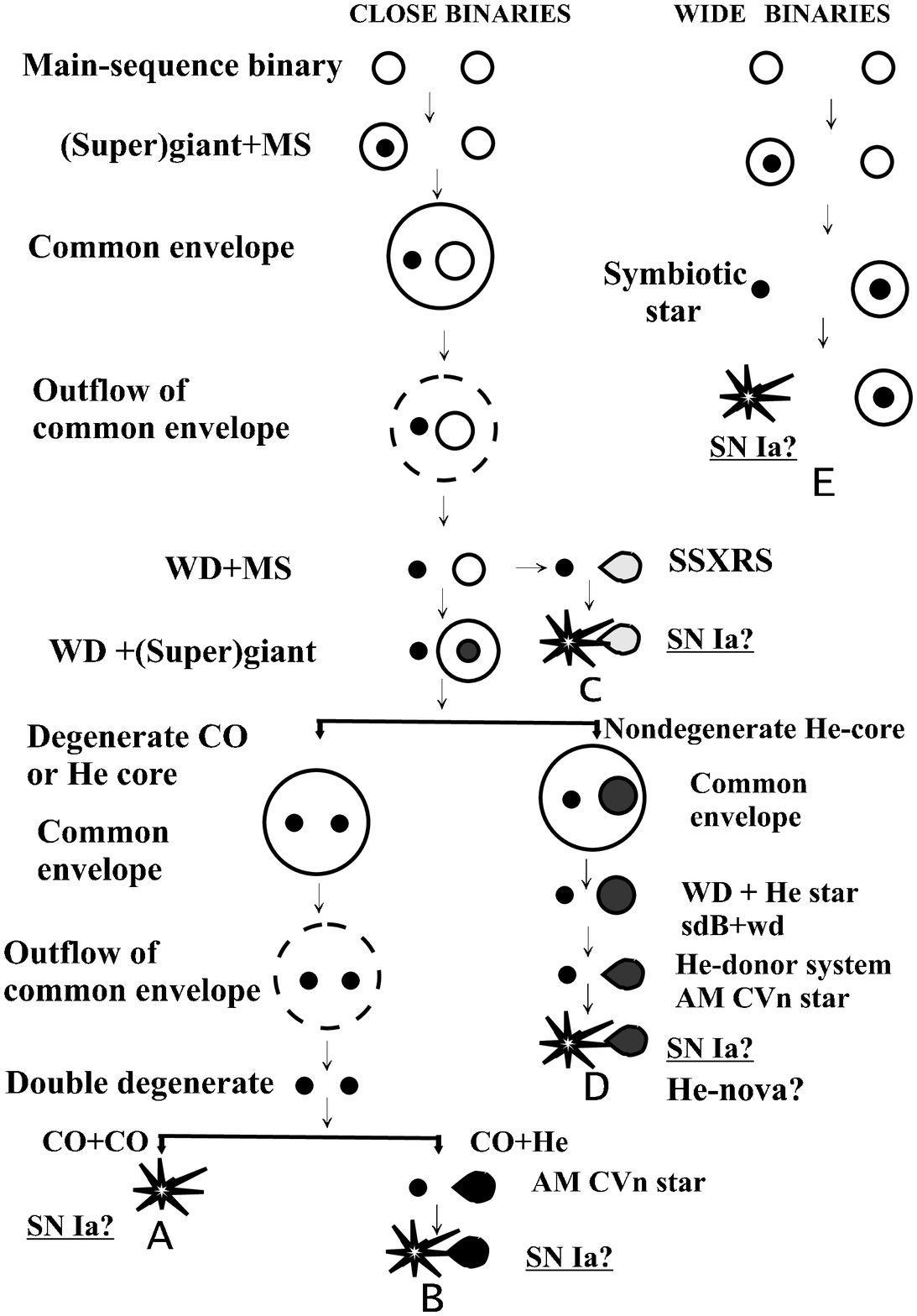}
   \caption{Main evolutionary channels resulting in candidate SN\,Ia systems.} 
\end{figure}

Figure \ref{fig:snflow} summarizes the main evolutionary scenarios that are currently considered to result in formation of candidate SN\,Ia systems. For completeness, the channel of symbiotic stars is shown, though, they are not ``close binaries'' strictly. A detailed review of the evolution to  SN\,Ia is recently published in \cite{y04}. Here we give  a brief outline.

DD-merger scenario (\textbf{A}) 
operates both in young ($\sim 10^9)$\,yr and old  ($\sim 10^{10})$\,yr populations \cite{ty02,y04} and may provide for our Galaxy the rate of SN\,Ia (\nusn) comparable to the one inferred from observations \cite{captur01}: $(4 \pm 2) \times 10^{-3}$\,\pyr.
The DD-scenario recently got strong support both from the theory and observations.
Population synthesis \citep{nyp+01} predicted that it may be necessary to study up to $\sim 1000$ field WD with $V \aplt 16\div17$ for finding a  proper merger candidate. This was done 
within SPY-project
and resulted in discovery of the first super-Chandrasekhar DD [$M_t=(1.46\pm0.12)\,\ms$] that will merge in  4 Gyr and two DD with
mass close to \mc\ 
that will merge in one to two Hubble times \cite{nap_spy03}. This means that formation of DD candidate SN\.Ia is quite possible.
On the ``theoretical'' side,
SPH calculations of DD-mergers with account for nuclear burning \citep{guerrero04}
have shown that in the  core-disc structure that forms,
in the region of interaction 
the temperature rises rapidly,
lifts degeneracy,
expands the matter and  
thus quenches thermonuclear flash. Flame does not propagate to the center, contrary to previously expected (e. g., \cite{mochko90}).
Stable accretion from the disk 
is possible. On the other hand,     
deposition of the angular momentum into the core 
spins-up  rotation of WD, causes 
 deformation of WD and AML by distorted configuration via GWR
and finally makes possible close-to-center explosive ignition of carbon \cite{piers+03a,piers+03b}.

Scenario \textbf{B} operates both in young and old populations, but 
gives a
minor contribution to \nusn\  since typical total masses of the systems are well below \mc.

``Single-degenerate'' (SD) scenario (\textbf{C}) is often considered as the most promising one. Population synthesis estimates lend to it from $\sim 10\%$ \cite{fty04} to $\sim 50\%$ \cite{hp04}  of the observed \nusn.
 The uncertainty in the estimates of \nusn\ from this channel stems from still 
poorly constrained efficiency of mass accumulation in the process of accretion that is accompanied by hydrogen and helium burning flashes \cite{it96symb,cassisi_etal98,piers+99}. The  estimates become  more favorable if
mass transfer may be stabilized by mass and momentum loss from the system. 
``Observational'' problem with this scenario is the absence of 
hydrogen in the spectra of SN\.Ia, while it is expected that
 $\sim 0.15$ \ms\ of H-rich matter may be stripped from the companion by the SN shell \cite{marietta00}.
Recently discovered SN\.Ia  2001ic and similar 1997cy with signatures of H in the spectra \citep{hamuy03} may belong to the so-called SN\,1.5 type or occur in  symbiotic systems \cite{cy04}.
 Also, under favorable for this scenario conditions, it seems to overproduce
supersoft X-ray sources \citep{fty04}. 
However, SD-scenario got recently  support from discovery of a possible companion
to Tycho SN \citep{rlapuente04}.

Scenario \textbf{D} may operate in populations where star formation have ceased no more than $\sim 1$\,Gyr ago and produce
SN-scale events at the rates that are comparable with the inferred Galactic SN\.Ia rate. But ``by construction'' of the model, the most rapidly moving products of explosions have to be He and Ni;
this is not observed.  The spectra produced by ELD are not compatible with observations of the overwhelming majority of SN\.Ia \cite{hk96}. On the theoretical side, it is possible that lifting effect of rotation that reduces effective gravity and degeneracy in the helium layer may prevent detonation  \citep{yoon_langer04}.

Scenario \textbf{E}  is the only way to produce SN\,Ia in a wide system, via accumulation of a He layer for ELD or of \mc\ by accretion of stellar wind matter in a symbiotic binary \cite{ty_symb76}. However, its efficiency is probably low ($\nusn \sim 10^{-6}$\,\pyr), because of typically low mass of WD and low $\dot{M}$ in symbiotic systems \cite{yltk95}.

\section{CONCLUSION}

For the conclusion, let us note that, despite evolution of close binary stars is, at least qualitatively, understood and for certain systems population synthesis provides even quantitative agreement with observations, there are some important problems to solve:

\noindent
$\bullet$ mass and angular momentum loss from the systems (including stellar winds), evolution in common envelopes;

\noindent
$\bullet$ processes during merger of stars, from main-sequence stars through relativistic objects; evolution of merger products; 

\noindent
$\bullet$ the role of rotation, angular momentum deposition during accretion 
and associated instabilities; 

\noindent
$\bullet$ mass limits between white dwarfs/neutron stars/black holes.

\vskip 0.3cm
The author acknowledges financial support from the LOC of the conference and from 
Russian Foundation for Basic Research
(grant no. 03-02-16254).




\end{document}